\documentclass[11pt]{article}

\usepackage{amsmath,amssymb}
\usepackage{rotating}
\usepackage{array}
\usepackage{epsfig}

\usepackage{amsmath}
\usepackage{graphicx}
\usepackage{latexsym}

\newcount\figureno     \figureno=0
\newdimen\figdim       \figdim=70mm
\def\figureinc{%
   \global\advance\figureno by 1%
}
\def\figcaption#1#2#3{\hbox to #2{\hss{\vbox{\hsize=#2 \parindent=0pt
        {\bf Figure \number\figureno#3 :\ }#1}}\hss}
}

\begin{document}
\baselineskip 100pt

{\large
\parskip.2in
\newcommand{\be}{\begin{equation}}
\newcommand{\ee}{\end{equation}}
\newcommand{\ben}{\begin{equation*}}
\newcommand{\een}{\end{equation*}}
\newcommand{\br}{\bar}
\newcommand{\fr}{\frac}
\newcommand{\lm}{\lambda}
\newcommand{\ra}{\rightarrow}
\newcommand{\al}{\alpha}
\newcommand{\bt}{\beta}
\newcommand{\z}{\zeta}
\newcommand{\pa}{\partial}
\newcommand{\hs}{\hspace{5mm}}
\newcommand{\up}{\upsilon}
\newcommand{\dg}{\dagger}
\newcommand{\sdil}{\ensuremath{\rlap{\raisebox{.15ex}{$\mskip
6.5mu\scriptstyle+ $}}\subset}}
\newcommand{\sdir}{\ensuremath{\rlap{\raisebox{.15ex}{$\mskip
6.5mu\scriptstyle+ $}}\supset}}
\newcommand{\vphi}{\vec{\varphi}}
\newcommand{\ve}{\varepsilon}
\newcommand{\acc}{\\[3mm]}
\newcommand{\dl}{\delta}
\def\tablecap#1{\vskip 3mm \centerline{#1}\vskip 5mm}
\def\p#1{\partial_#1}
\newcommand{\pd}[2]{\frac{\partial #1}{\partial #2}}
\newcommand{\pdn}[3]{\frac{\partial #1^{#3}}{\partial #2^{#3}}}
\def\DP#1#2{D_{#1}\varphi^{#2}}
\def\dP#1#2{\partial_{#1}\varphi^{#2}}
\def\xh{\hat x}
\newcommand{\Ref}[1]{(\ref{#1})}

\def\mod#1{ \vert #1 \vert }
\def\chapter#1{\hbox{Introduction.}}
\def\Sin{\hbox{sin}}
\def\Cos{\hbox{cos}}
\def\Exp{\hbox{exp}}
\def\Ln{\hbox{ln}}
\def\Tan{\hbox{tan}}
\def\Cot{\hbox{cot}}
\def\Sinh{\hbox{sinh}}
\def\Cosh{\hbox{cosh}}
\def\Tanh{\hbox{tanh}}
\def\Asin{\hbox{asin}}
\def\Acos{\hbox{acos}}
\def\Atan{\hbox{atan}}
\def\Asinh{\hbox{asinh}}
\def\Acosh{\hbox{acosh}}
\def\Atanh{\hbox{atanh}}
\def\frac#1#2{{\textstyle{#1\over #2}}}

\def\ph{\varphi_{m,n}}
\def\phl{\varphi_{m-1,n}}
\def\phr{\varphi_{m+1,n}}
\def\varphil{\varphi_{m-1,n}}
\def\varphir{\varphi_{m+1,n}}
\def\varphit{\varphi_{m,n+1}}
\def\varphib{\varphi_{m,n-1}}
\def\pht{\varphi_{m,n+1}}
\def\phb{\varphi_{m,n-1}}
\def\phbl{\varphi_{m-1,n-1}}
\def\phbr{\varphi_{m+1,n-1}}
\def\phtl{\varphi_{m-1,n+1}}
\def\phtr{\varphi_{m+1,n+1}}
\def\u{u_{m,n}}
\def\ul{u_{m-1,n}}
\def\ur{u_{m+1,n}}
\def\ut{u_{m,n+1}}
\def\ub{u_{m,n-1}}
\def\utr{u_{m+1,n+1}}
\def\ubl{u_{m-1,n-1}}
\def\utl{u_{m-1,n+1}}
\def\ubr{u_{m+1,n-1}}
\def\v{v_{m,n}}
\def\vl{v_{m-1,n}}
\def\vr{v_{m+1,n}}
\def\vt{v_{m,n+1}}
\def\vb{v_{m,n-1}}
\def\vtr{v_{m+1,n+1}}
\def\vbl{v_{m-1,n-1}}
\def\vtl{v_{m-1,n+1}}
\def\vbr{v_{m+1,n-1}}

\def\U{U_{m,n}}
\def\Ul{U_{m-1,n}}
\def\Ur{U_{m+1,n}}
\def\Ut{U_{m,n+1}}
\def\Ub{U_{m,n-1}}
\def\Utr{U_{m+1,n+1}}
\def\Ubl{U_{m-1,n-1}}
\def\Utl{U_{m-1,n+1}}
\def\Ubr{U_{m+1,n-1}}
\def\V{V_{m,n}}
\def\Vl{V_{m-1,n}}
\def\Vr{V_{m+1,n}}
\def\Vt{V_{m,n+1}}
\def\Vb{V_{m,n-1}}
\def\Vtr{V_{m+1,n+1}}
\def\Vbl{V_{m-1,n-1}}
\def\Vtl{V_{m-1,n+1}}
\def\Vbr{V_{m+1,n-1}}

\newcommand{\ie}{{\it i.e.}}
\newcommand{\cmod}[1]{ \vert #1 \vert ^2 }
\newcommand{\cmodn}[2]{ \vert #1 \vert ^{#2} }
\newcommand{\nhat}{\mbox{\boldmath$\hat n$}}
\nopagebreak[3]
\bigskip

\title{ \bf Fermionic extension of the scalar Born-Infeld equation and its relation to the supersymmetric Chaplygin gas}
\vskip 1cm

\bigskip
\author{
A J Hariton\thanks{email address: hariton@lns.mit.edu}\\\\
Massachusetts Institute of Technology\\ Center for Theoretical Physics NE25-4061 \\ 77 Massachusetts Avenue\\ Cambridge, MA 02139, United States\\\\ Tel: 617-253-7195\\ Fax: 617-253-8674} \date{}

\maketitle

\begin{abstract}

In this article, we present a fermionic extension of the scalar Born-Infeld equation, which has been derived from the Nambu-Goto superstring action in $(2+1)$ dimensions through the Cartesian parameterization. It is demonstrated that in the relativistic limit where $c\rightarrow\infty$, the fermionic Born-Infeld model reduces to the supersymmetric Chaplygin gas model in one spatial dimension. However, the supersymmetry itself is not preserved.

\end{abstract}


\vspace{3mm}



\vspace{3mm}

\begin{center}MIT-CTP-3746\end{center}

\newpage

\section{Introduction} 

The theory of nonlinear electrodynamics of Born and Infeld had its origins in the work of Mie \cite{Mie}, who attempted to formulate a unitarian theory of electromagnetics through a nonlinear generalization of Maxwell's equations. However, Mie's theory possessed the disadvantage that the electromagnetic potentials acquired a direct physical significance, leading to the loss of gauge invariance. Born and Infeld postulated a nonlinear, gauge-invariant, relativistic theory \cite{Born1,Born2} described by the Lagrangian
\begin{equation}
L_{BI}\,=\,b^2\left(1\,-\,\sqrt{1 -  \left(\mathbf{E}^2-\mathbf{B}^2\right)/{b^2} - \left(\mathbf{E}\cdot\mathbf{B}\right)^2/{b^4}}\right)\mbox{,}
\label{BI1rowr}
\end{equation}
where $\mathbf{E}$ and $\mathbf{B}$ represent the electric and magnetic fields respectively, and $b$ is a parameter having the dimension of the electromagnetic field \cite{Born2,Bialynicki}. The laws of propagation of photons and charged particles were studied by Boillat \cite{Boillat}, and it was shown that the Born-Infeld theory leads to propagation without birefringence or shock waves. More generally, equations derived from an action principle involving a square root similar to that of (\ref{BI1rowr}) are referred to as Born-Infeld type equations in the literature on the subject. Such generalized Born-Infeld Lagrangians have appeared, for example, in string theories involving scalar (dilaton) fields and gravity \cite{Kerner3,Gibbons}. More recently, a generalization of the Born-Infeld Lagrangian for non-abelian gauge theory was proposed \cite{Kerner2} and adapted to the noncommutative geometry of matrix valued functions on a manifold \cite{Kerner1}.

A few years ago, a series of lectures was given by Professor R Jackiw in which the subject of fluid mechanics was examined from an entirely new perspective \cite{Jackiw}. A number of topics was covered, including the equations of motion, conservation laws, and a general description of the properties of classical fluids. Of particular interest were the the Galileo-invariant Chaplygin gas model, which describes an isentropic, non-dissipative fluid with polytropic pressure \cite{Chaplygin,Landau}, as well as the Poincar\'{e}-invariant Born-Infeld scalar model used to describe the interaction of two plane waves \cite{Barbashov1,Barbashov2}. In particular, it was shown that both models  devolve from the parameterization-invariant Nambu-Goto action for a $d$-brane evolving in $(d+1,1)$-dimensional space-time
\begin{equation}
I_{NG} = \int\sqrt{(-1)^d\det{\left({\partial X^{\mu}\over \partial\phi^i}{\partial X_{\mu}\over \partial\phi^j}\right)}}d\phi^0d\phi^1\ldots\phi^d\mbox{.}
\label{intro1}
\end{equation}
Here, $\phi^0,\phi^1,\ldots,\phi^d$ are the world-volume variables which parameterize the extended object, while the variables $X^0,X^1,\ldots,X^d,X^{d+1}$ describe the target space-time. Two distinct choices of parameterization of the action (\ref{intro1}) lead respectively to the Chaplygin and Born-Infeld models. In both cases, $(X^1,\ldots,X^d)$ is chosen to coincide with $(\phi^1,\ldots,\phi^d)$, and they are called $\mathbf{r}$ (a $d$-dimensional position vector). For the light-cone parameterization, the quantity $(X^0+X^{d+1})/\sqrt{2}$ is renamed $t$ and identified with $\sqrt{2\lambda}\phi^0$, while $(X^0-X^{d+1})/\sqrt{2}$ is identified with $\theta$, a function of $t$ and $\mathbf{r}$. The Nambu-Goto action (\ref{intro1}) reduces to the Chaplygin gas action
\begin{equation}
I_{\lambda} = -2\sqrt{\lambda}\int\sqrt{\partial_t\theta + \frac{1}{2}\left(\mathbf{\nabla}\theta\right)^2} dtd\mathbf{r}\mbox{.}
\label{intro2}
\end{equation}
%
For the Cartesian parameterization, $X^0$ is renamed $ct$ and identified with $\phi^0$, while $X^{d+1}$ is renamed $\theta/c$, a function of $t$ and $\mathbf{r}$. The Nambu-Goto action (\ref{intro1}) then coincides with the scalar Born-Infeld action
\begin{equation}
I_{a} = -a\int\sqrt{c^2-\left(\partial_{\mu}\theta\right)^2} dtd\mathbf{r}\mbox{.}
\label{intro3}
\end{equation}
It should also be noted that at the nonrelativistic limit where $c\rightarrow\infty$, the Born-Infeld action (\ref{intro3}) reduces to the Chaplygin action (\ref{intro2}) provided that the nonrelativistic function $\theta_{NR}$ in (\ref{intro2}) is extracted from the relativistic function $\theta_R$ in (\ref{intro3}) by the relation
\begin{equation}
\theta_R = -c^2t + \theta_{NR}\mbox{.}
\label{intro4}
\end{equation}

Moreover, it was shown that fluid mechanics can be enhanced through the addition of fermionic (anticommuting Grassmannian) degrees of freedom. In particular, supersymmetric generalizations were formulated for the Chaplygin gas in one and two spatial dimensions \cite{Polychronakos,Bergner}. In the case of one spatial dimension, the resulting Lagrangian density reads
\begin{equation}
{\cal L} = -\sqrt{2\partial_t\theta - \psi\partial_t\psi + (\partial_x\theta - \frac{1}{2}\psi\partial\psi)^2} + \frac{1}{2}\psi\partial_x\psi\mbox{,}
\label{intro5}
\end{equation}
where $\theta$ is the standard bosonic scalar field of the Chaplygin gas equation and $\psi$ is a real fermionic Grassmann variable.

 It was also demonstrated how the equations for the supersymmetric Chaplygin fluid devolve from a supermembrane Lagrangian, through the light-cone parameterization mentioned above. The question arises as to whether an analogous supersymmetric extension can be formulated for the scalar Born-Infeld model by applying the Cartesian parameterization to the supermembrane Lagrangian. In this paper, we attemp to answer this question for the case of one spatial dimension.

This paper is organized as follows. In Section 2, we parameterize the Nambu-Goto action for a superstring using the Cartesian parameterization and evaluate the resulting Lagrangian. It is demonstrated that this Lagrangian is indeed a Lorentz-invariant fermionic extension of the one for the scalar Born-Infeld equation. Furthermore, in the nonrelativistic limit where $c\rightarrow\infty$, the theory reduces to that of the supersymmetric Chaplygin Lagrangian in one spatial dimension (\ref{intro5}). In section 3, we discuss the Hamiltonian and the canonical form of the Lagrangian and derive the equations of motion. Finally in Section 4, we discuss a supersymmetry of the Nambu-Goto superstring theory and the fact that this supersymmetry is not carried over when we go to the Cartesian parameterization.

\section{Fermionic extension of the scalar Born-Infeld model}

We begin with the Nambu-Goto action for a superstring in a $(2+1)$ dimensional target space-time
\begin{equation}
I = -\int \left(\sqrt{g} - i\epsilon^{ij}\partial_iX^{\mu}\bar{\psi}\gamma_{\mu}\partial_j\psi\right)d\phi^0d\phi^1\mbox{,}
\label{eq1}
\end{equation}
where
\begin{equation}
g = -\det{\left[\left(\partial_iX^{\mu}-i\bar{\psi}\gamma^{\mu}\partial_i\psi\right)\left(\partial_jX^{\nu}-i\bar{\psi}\gamma^{\nu}\partial_j\psi\right)\eta_{\mu\nu}\right]}\mbox{.}
\label{eq1A}
\end{equation}
Henceforth, $\mu$ and $\nu$ are indices running over $0,1,2$, which represent the target space-time, and $i,j$ are the worldsheet indices $0,1$. We use the symbol $\partial_i$ to denote derivation by $\phi^i$ and the $\gamma^{\mu}$ matrices are defined as
\begin{equation}
\gamma^0 = \sigma^1 = \begin{pmatrix} 0 & 1 \\ 1 & 0\end{pmatrix}\mbox{,}\qquad \gamma^1 = i\sigma^2 = \begin{pmatrix} 0 & 1 \\ -1 & 0\end{pmatrix}\mbox{,}\qquad
\gamma^2 = i\sigma^3 = \begin{pmatrix} i & 0 \\ 0 & -i\end{pmatrix}\mbox{,}
\end{equation}
\vspace{2mm}
\begin{equation}
\gamma^5 = \gamma^0\gamma^1 = \begin{pmatrix} -1 & 0 \\ 0 & 1\end{pmatrix} = i\gamma^2\mbox{.}
\label{eq2}
\end{equation}
Here, $\psi$ is a real two-component spinor $\psi = (u,v)$ whose components are odd (fermionic) Grassmann variables. For the purpose of deriving the generalized fermionic Born-Infeld action, we chose the following Cartesian parameterization
\begin{equation}
X^0 = ct\mbox{,}\qquad X^1 = x\mbox{,}\qquad X^2 = {1\over c}\theta(t,x)\mbox{,}\qquad \phi^0 = ct\mbox{,}\qquad \phi^1 = x\mbox{.}
\label{eq3}
\end{equation}
The components of the matrix $g_{ij}$ in equation (\ref{eq1A}) are then given by
\begin{equation}
\begin{split}
g_{ij} &= \eta_{ij} - {1\over c^2}(\partial_i\theta)(\partial_j\theta) - i\bar{\psi}\gamma_i\partial_j\psi - i\bar{\psi}\gamma_j\partial_i\psi + {1\over c}i(\partial_i\theta)\bar{\psi}\gamma^2\partial_j\psi\\ & \mbox{  } + {1\over c}i(\partial_j\theta)\bar{\psi}\gamma^2\partial_i\psi + 3\bar{\psi}\partial_i\psi\bar{\psi}\partial_j\psi\mbox{,}
\end{split}
\label{eq4}
\end{equation}
where we have used the identity
\begin{equation}
\left(\sigma^{\mu}\right)_{ij}\left(\sigma^{\mu}\right)_{kl} = 2\delta_{il}\delta_{jk} - \delta_{ij}\delta_{kl}\qquad (\mu = 0,1,2)\mbox{.}
\label{eq5}
\end{equation}
The supplementary term in the action (\ref{eq1}) is
\begin{equation}
-i\epsilon^{ij}\bar{\psi}\gamma_i\partial_j\psi + {1\over c}i\epsilon^{ij}(\partial_i\theta)\bar{\psi}\gamma^2\partial_j\psi\mbox{.}
\label{eq6}
\end{equation}
The fermionic gauge choice
\begin{equation}
(1+\gamma^5)\psi = 0\mbox{,}
\label{eq7}
\end{equation}
reduces the spinor $\psi$ to an one-component odd Grassmann field, which we renormalize to ${1\over \sqrt{2ic}}u$, with real $u$. The components of the matrix $g$ then become
\begin{equation}
\begin{split}
g_{00} &= 1 - {1\over c^4}(\theta_t)^2 - {1\over c^2}uu_t\mbox{,}\\
g_{01} = g_{10} &= -{1\over c^3}\theta_x\theta_t - {1\over 2c}uu_x - {1\over 2c^2}uu_t\mbox{,}\\
g_{11} &= -1 -{1\over c^2}(\theta_x)^2 - {1\over c}uu_x\mbox{,}
\end{split}
\label{eq8}
\end{equation}
so that $g = g_{01}^2 - g_{00}g_{11}$ is
\begin{equation}
\begin{split}
g &= 1 - {1\over c^4}(\theta_t)^2 + {1\over c^2}(\theta_x)^2 + {1\over c}uu_x - {1\over c^2}uu_t - {1\over c^5}(\theta_t)^2uu_x - {1\over c^4}(\theta_x)^2uu_t\\ &+ {1\over c^4}\theta_x\theta_tuu_x + {1\over c^5}\theta_x\theta_tuu_t\mbox{.}
\end{split}
\label{eq9}
\end{equation}
The supplementary term (\ref{eq6}) becomes $-{1\over 2c}uu_x + {1\over 2c^2}uu_t$. Multiplying by a factor of $c$, the Lagrangian density of the action (\ref{eq1}) reduces to the following generalization of the scalar Born-Infeld Lagrangian
\begin{equation}
\begin{split}
{\mathcal L} = &-\Bigg{(}c^2 - {1\over c^2}(\theta_t)^2 + (\theta_x)^2 + cuu_x - uu_t - {1\over c^3}(\theta_t)^2uu_x - {1\over c^2}(\theta_x)^2uu_t\\ & + {1\over c^2}\theta_x\theta_tuu_x + {1\over c^3}\theta_x\theta_tuu_t\Bigg{)}^{1/2} + {1\over 2}uu_x - {1\over 2c}uu_t\mbox{.}
\end{split}
\label{eq11}
\end{equation}
It is evident that in the case where $u\rightarrow 0$, the Lagrangian (\ref{eq11}) reduces to the standard (non-fermionic) Lagrangian density for the Born-Infeld scalar model. In addition, the nonrelativistic limit of this theory is the supersymmetric Chaplygin model in $(1+1)$ dimensions formulated by Bergner and Jackiw. Indeed, if we extract from the relativistic field $\theta_R$ 
the nonrelativistic field $\theta_{NR}$ 
in such a way that
\begin{equation}
\theta_R = -c^2t + \theta_{NR}\mbox{,}
\label{eq12}
\end{equation}
and take the limit as $c\rightarrow\infty$, we obtain the Lagrangian
\begin{equation}
{\mathcal L} = -\sqrt{2\theta_t - uu_t + (\theta_x)^2 - \theta_xuu_x} + {1\over 2}uu_x\mbox{,}
\label{eq13}
\end{equation}
which is precisely the Lagrangian described in \cite{Bergner}.

Although it is not evident that the Lagrangian (\ref{eq11}) is Lorentz invariant, this can be demonstrated in the following way. Define the Lorentz operator $L$ to be
\begin{equation}
L = ct\partial_x + {1\over c}x\partial_t\mbox{,}
\label{eq14}
\end{equation}
and consider the variation of $L$ on the fields $\theta$ and $u$:
\begin{equation}
\delta\theta = L\theta\mbox{,}\qquad \delta u = Lu + {1\over 2}u\mbox{.}
\label{eq15}
\end{equation}
%
A straightforward calculation then determines that $\delta{\cal L} = L{\cal L}$ so that the theory is indeed Lorentz invariant. The Lagrangian (\ref{eq11}) can also be written in the form
\begin{equation}
\begin{split}
{\cal L} = &-\left(c^2 - (\partial_{\mu}\theta)^2 - cun^i\partial_iu - {1\over c}n^i\partial_i\theta u\epsilon^{jk}\partial_{j}\theta\partial_{k}u\right)^{1/2}\\ & - {1\over 2}un^i\partial_iu\mbox{,}
\end{split}
\label{eq17}
\end{equation}
where $n^i$ is the light-like vector
\begin{equation}
n^i = (1,-1)\mbox{,}
\label{eq17AA}
\end{equation}
whose inclusion does not disrupt the Lorentz invariance of ${\cal L}$.

\section{Canonical formulation}

In the case of the supersymmetric Chaplygin gas, an alternative formulation of the Lagrangian was used, which included not only the fields $\theta$ and $u$, but also an additional bosonic field $\rho$, which played the role of fluid density. This density corresponded to the canonical momentum of the scalar field $\theta$, taken with respect to the time $t$. Although it is very difficult to proceed in the same way in the case of the Born-Infeld Lagrangian, the problem can be made easier by making use of the light-cone basis:
\begin{equation}
\theta_{\pm} = {1\over \sqrt{2}}\left({1\over c}\theta_t \pm \theta_x\right)\mbox{,}\qquad 
u_{\pm} = {1\over \sqrt{2}}\left({1\over c}u_t \pm u_x\right)\mbox{.}
\label{eq18}
\end{equation}
The Lagrangian (\ref{eq11}) can then be re-written as
\begin{equation}
{\cal L} = -\sqrt{c^2 - 2\theta_-\theta_+ - \sqrt{2}cuu_- - {\sqrt{2}\over c}\theta_-^2uu_+ + {\sqrt{2}\over c}\theta_+\theta_-uu_-} - {\sqrt{2}\over 2}uu_-\mbox{,}
\label{eq19}
\end{equation}
and we determine the canonical momentum of $\theta$ with respect to the $(+)$ coordinate to be
\begin{equation}
\Pi = {\partial{\cal L}\over \partial\theta_+} = {\theta_- - {1\over \sqrt{2}c}\theta_-uu_-\over \sqrt{c^2 - 2\theta_-\theta_+ - \sqrt{2}cuu_- - {\sqrt{2}\over c}\theta_-^2uu_+ + {\sqrt{2}\over c}\theta_+\theta_-uu_-}}\mbox{.}
\label{eq20}
\end{equation}
The derivative $\theta_+$ can therefore be determined in terms of $\Pi$:
\begin{equation}
\theta_+ = {\theta_-\left(1-{\sqrt{2}\over c}uu_-\right)\over 2\left({1\over \sqrt{2}c}uu_--1\right)\Pi^2} + {{\sqrt{2}\over c}\theta_-^2uu_++{\sqrt{2}}cuu_--c^2\over 2\theta_-\left({1\over \sqrt{2}c}uu_--1\right)}\mbox{,}
\label{eq21}
\end{equation}
and the Hamiltonian of the system evaluated as
\begin{equation}
{\cal H} = \Pi\theta_+-{\cal L}\mbox{.}
\label{eq22}
\end{equation}
The rational terms in $u$ may be simplified by a series expansion which, due to the Grassmannian nature of $u$, will vanish for terms of degree $2$ or higher. The resulting expression for ${\cal H}$ is
\begin{equation}
{\cal H} = {\theta_-\over 2\Pi}\left(1-{1\over \sqrt{2}c}uu_-\right) + {\Pi\over 2\theta_-}\left(c^2-{1\over \sqrt{2}}cuu_--{\sqrt{2}\over c}\theta_-^2uu_+\right) + {1\over \sqrt{2}}uu_-\mbox{.}
\label{eq23}
\end{equation}
The Lagrangian, written in terms of the variables $\theta$, $\Pi$ and $u$ is therefore
\begin{equation}
{\cal L} = \Pi\theta_+ + {1\over \sqrt{2}c}\Pi\theta_-uu_+ + \left({\Pi c^2\over 2\theta_-}+{\theta_-\over 2\Pi}\right)\left({1\over \sqrt{2}c}uu_--1\right) - {1\over \sqrt{2}}uu_-\mbox{.}
\label{eq24}
\end{equation}
The equations of motion are
\begin{equation}
\theta_+ + {1\over\sqrt{2}c}\theta_-uu_+ + \left({c^2\over 2\theta_-}-{\theta_-\over 2\Pi^2}\right)\left({1\over\sqrt{2}c}uu_--1\right) = 0\mbox{,}
\label{eq25}
\end{equation}
and 
\begin{equation}
\partial_+\Pi + \partial_-\left({1\over\sqrt{2}c}\Pi uu_+ + \left({1\over 2\Pi}-{\Pi c^2\over 2\theta_-^2}\right)\left({1\over \sqrt{2}c}uu_--1\right)\right) = 0\mbox{.}
\label{eq26}
\end{equation}

\section{Supersymmetry}
 
The action of the supersymmetric extension of the Chaplygin gas in one spatial dimension (\ref{intro5}) is invariant under the supersymmetry
\begin{equation}
\delta\theta = \frac{1}{2}\eta\psi\mbox{,}\qquad \delta\psi = \eta\mbox{,}
\label{susskind1}
\end{equation}
and this is related to the fact that the Nambu-Goto action for a superstring (\ref{eq1}) is itself invariant under the transformation
\begin{equation}
\delta X^{\mu} = i\bar{\eta}\gamma^{\mu}\psi\mbox{,}\qquad \delta\psi = \eta\mbox{.}
\label{susskind2}
\end{equation}
Indeed, the supersymmetric transformation (\ref{susskind1}) is obtained readily from (\ref{susskind2}) when we go to the light-cone parameterization.
It is therefore natural to ask whether the transformation (\ref{susskind2}) could be carried over to an equivalent supersymmetry of the fermionic Born-Infeld action (\ref{eq11}) when we go to the Cartesian parameterization.

It is evident that in the general (non-parameterized) case of the superstring, the quantity of the form $A^{\mu}_i = \partial_iX^{\mu}-i\bar{\psi}\gamma^{\mu}\partial_i\psi$ is preserved by the transformation (\ref{susskind2}) for all values of $\mu$ and all $i$. However, when the Cartesian parameterization is applied, for instance $\delta A^0_0$ becomes 
\begin{equation}
\delta A^0_0 = \delta \left(\partial_0X^{0}-i\bar{\psi}\gamma^{0}\partial_0\psi\right)\ \propto\  -\eta\partial_0u \neq 0\mbox{,}
\label{susskind3}
\end{equation}
which loses the supersymmetry. (The proportionality is determined by the normalization.) It remains an open question as to whether a supersymmetric transformation exists for this specific theory. On the other hand, it has been demonstrated that one can construct supersymmetric extensions of the scalar Born-Infeld equation using a superspace formalism \cite{Hariton}. These differ from the present one in that they do not appear to descend from the Nambu-Goto action, nor do they reduce to the Chaplygin gas in a nonrelativistic limit.

\subsection*{Acknowledgements}

The author would like to thank Professor R. Jackiw for helpful discussions and useful information on related matters. This work is supported in part by funds provided by the U.S. Department of Energy (D.O.E.) under cooperative research agreement DEFG02-05ER41360.

{}

\label{lastpage}
\end{document}